\documentclass{jpp}
\usepackage{graphicx}

\usepackage[utf8]{inputenc}
\usepackage[T1]{fontenc}
\usepackage{amsmath}

\usepackage{color}

\newcommand{\phib}{\overline{\delta\phi}}
\newcommand{\bn}{\begin{equation}}
\newcommand{\bea}{\begin{eqnarray*}}
\newcommand{\eea}{\end{eqnarray*}}
\newcommand{\en}{\end{equation}}
\newcommand{\lang}{\left\langle}
\newcommand{\rang}{\right\rangle}

\newcommand{\vth}{v_{th}}
\newcommand{\vpar}{v_\parallel}
\newcommand{\vperp}{v_\perp}

\newcommand{\osT}{\eta\omega_\ast}
\newcommand{\od}{\omega_d}
\newcommand{\odt}{\tilde{\omega}_d}

\providecommand\bkappa{\boldsymbol{\kappa}}

\newcommand{\Ephi}{E}
\newcommand{\Dphi}{K}

\shorttitle{Energetic bounds.  Part III}
\shortauthor{G. G. Plunk and P. Helander}

\title{Energetic bounds on gyrokinetic instabilities. Part III. Generalized free energy.}

\author{G. G. Plunk\aff{1}
  \corresp{\email{gplunk@ipp.mpg.de}},
  \and P. Helander\aff{1}}

\affiliation{\aff{1}Max-Planck-Institut für Plasmaphysik, 17491 Greifswald, Germany}

\begin{document}

\maketitle

\begin{abstract}
Free energy, widely used as a measure of turbulence intensity in weakly collisional plasmas, has been recently found to be a suitable basis to describe both linear and nonlinear growth in a wide class gyrokinetic systems.  The simplicity afforded by this approach is accompanied by some drawbacks, notably the lack of any explicit treatment of wave-particle effects, which makes the theory unable to describe things like stability thresholds or dependence on the geometry of the background magnetic field.  As a step toward overcoming these limitations, we propose an extension of the theory based on a generalization of free energy.  With this it is demonstrated that resonance effects are recovered, and the bounds on growth are significantly reduced.  The simplicity and efficient computation of the associated ``optimal'' growth rates makes the theory potentially applicable to stellarator optimization.
\end{abstract}

\section{Introduction}

This is the third paper in a series \citep{helander_plunk_2022, plunk_helander_2022}, in which we develop a linear and nonlinear stability theory based on gyrokinetic energy balance.  The last two papers used Helmholtz free energy, and introduced the concept of optimal mode growth for fully electromagnetic gyrokinetics.  The present paper proposes a generalized energetic measure of fluctuations, allowing the inclusion of additional instability mechanisms.  We do this first for a simple case, namely the electrostatic limit (low plasma $\beta$) with only one kinetic species (ions), with the electrons being treated adiabatically.  These simplifications limit the application to ion-temperature-gradient (ITG) driven turbulence, though the central result of the paper is capable of treating completely general details of the magnetic geometry.

Free energy is a useful concept for understanding nonlinear and linear aspects of plasma turbulence.  At the level of linear instabilities it is common to speak of a source of free energy that drives modes.  Indeed, without a source of free energy, provided by background plasma gradients (density, temperature, flows), there can be no linear instabilities (nor can there be subcritical turbulence \cite{landreman_plunk_dorland_2015, plunk_helander_2022}).  However, there is usually another ingredient that arises in the detailed analysis of normal linear instabilities, namely the wave-particle resonance.  In gyrokinetic theory, this involves parallel motion (along the magnetic field) and magnetic drift, and the resonance is physically linked to the work that the electrostatic field performs on gyrocenter motion.  However, the terms needed to capture this do not contribute to free energy balance, and the influence of resonance therefore cannot be accounted for by the optimal modes that we introduced in our previous works.

In this work we propose a new measure of gyrokinetic fluctuations, a generalization of the concept of free energy, that incorporates the resonance mechanism, and, via the magnetic drift, the full details of the background magnetic geometry.  We demonstrate the existence of a class of quadratic measures closely related to Helmholtz free energy that behave as positive-definite norms for fluctuations in the distribution function.  The corresponding energy balance equation is then used to derive a theory of optimal modes that most efficiently extract this energy from its source.  The growth rate of these optimal modes provides a rigorous upper bound on the growth rate of linear instabilities, and this bound is shown to be lower than that obtained previously from Helmholtz free energy.  By studying some simple limits, we show that we recover some expected behavior of both the slab and toroidal branches of the ITG mode.

\section{Definitions and gyrokinetic energy balance}
We follow closely the conventions of the first paper in this series \citep{helander_plunk_2022}, focusing on local gyrokinetic theory in the geometry of a flux tube, whereby fluctuations in the distribution function may be considered small, and periodic in the coordinates perpendicular to the field line.  Essential definitions are summarized in what follows, but more detail and background can be found in sections 2 and 3 of \citet{helander_plunk_2022} (henceforth also called `Paper I').

The ion gyrokinetic equation in the electrostatic limit is written

\begin{equation}
    \frac{\partial g_{\bf k}}{\partial t} + v_{\|} \frac{\partial g_{\bf k}}{\partial l} + i \odt g_{\bf k} + \frac{1}{B^2} \sum_{{\bf k}'} {\bf B} \cdot ({\bf k} \times {\bf k}')  \phib_{{\bf k}'} g_{{\bf k} - {\bf k}'} = \frac{e_i F_{0}}{T_i} \left( \frac{\partial}{\partial t} + i \omega_{\ast}^T \right) \phib_{\bf k}, \label{gk-eqn}
\end{equation}
where $g$ is the gyro-center dependent part of the perturbed ion distribution function, {\em i.e.} $f_i = \left( 1 - e_i \delta \phi({\bf r})/T_i \right)F_{i0}  + g({\bf R}, E_i, \mu_i, t).$  Its phase space variables are the energy $E_a = m_a v^2 / 2 + e_a \Phi(\psi)$ and the magnetic moment $\mu_a = m_a v_\perp^2 / (2B)$, and the perpendicular wavenumber is ${\bf k} = {\bf k}_\perp = k_\psi \bnabla \psi + k_\alpha \bnabla \alpha$ with $k_\psi$ and $k_\alpha$ independent of the arc length $l$ along the magnetic field, and $\psi$ and $\alpha$ defined via ${\bf B} = B {\bf b} = \nabla \psi \times \nabla \alpha$.  We neglect collisions here\footnote{We do not retain collisions, since we will not be able to fix the sign of its contribution in our later analysis.}, and used the simplified notation $g_{\bf k} = g_{i, {\bf k}}$, and $\omega_{\ast} = \omega_{\ast i}$, {\em etc} because the adiabatic approximation $g_{e, {\bf k}} = 0$ is assumed throughout; note that it is not necessary here to include the customary correction for the zonal component ({\em i.e.} for $k_\alpha = 0$; see \citet{dorland-hammett-POP-1993}) because the growth of this component is zero due to the fact that it has no source of free energy (see below).  We will also assume $k \rho_i \sim 1$, implying $k\rho_e \ll 1$.

As derived for the general case in Section 3 of Paper I, the gyrokinetic free energy balance equation obtained in this limit reads

\begin{equation}
    \frac{d}{d t}\sum_{\bf k} H = 2 \sum_{\bf k} D,\label{energy-balance-H}
\end{equation}
where the drive term $D$ is

\begin{equation}
    D({\bf k}, t) = {\rm Im} \; e_i \lang \int g_{\bf k} \omega_{\ast}^T \phib^\ast_{\bf k} d^3v \rang, 
	\label{D-eqn}
\end{equation}
and the free energy, expressed in terms of the gyrocenter distribution function

\begin{equation}
    H({\bf k},t)  = \lang  T_i \int \frac{|g_{\bf k}|^2}{F_{i0}} d^3v - \sum_a \frac{n_a e_a^2}{T_a} |\delta \phi_{\bf k}|^2 \rang,\label{H-eqn}
\end{equation}
where the space average is defined as (extensions are discussed in section 3 of Paper I)

\begin{equation}
    \lang \cdots \rang = \lim_{L\rightarrow \infty} \int_{-L}^L (\cdots ) \frac{dl}{B} \bigg\slash
	\int_{-L}^L \frac{dl}{B}.
\end{equation}
The diamagnetic frequencies are

	$$ \omega_{\ast a} = \frac{k_\alpha T_a}{e_a} \frac{d \ln n_a}{d \psi}, $$
	$$ \omega_{\ast a}^T = \omega_{\ast a} 
	\left[1 + \eta_a \left( \frac{m_a v^2}{2 T_a} - \frac{3}{2} \right)\right], $$
and the magnetic drift frequency  is 

$$ \odt = {\bf k} \cdot {\bf v}_d, $$
where the magnetic drift velocity is ${\bf v}_d = \hat{\bf b}\times((v_{\perp}^2/2)\bnabla \ln B  + v_\|^2\bkappa)/\Omega_i$, $\bkappa = \hat{\bf b}\cdot\bnabla\hat{\bf b}$, and $\Omega_a = e_a B / m_a$ is the gyrofrequency.  Noting $\hat{\bf b}\times(\bnabla\ln B -\bkappa - \mu_0/B^2 \bnabla p) = 0$, we can assume $\hat{\bf b}\times\bnabla\ln B \approx \hat{\bf b}\times \bkappa$ in the appropriate limit of low plasma $\beta$, which allows us to separate the drift frequency into velocity-dependent and space-dependent factors following \cite{plunk-POP-2014}\footnote{Actually, there is spatial dependence in both $\vperp$ and $\vpar$, since these are not the proper gyrokinetic phase-space variables, but a separation like this is useful to make contact with known limits from gyrokinetic theory of the ITG mode.}:

\begin{equation}
    \odt = \od(l) \left[ \frac{\vperp^2}{2\vth^2} + \frac{\vpar^2}{\vth^2} \right].
\end{equation}
The gyro-averaged electrostatic potential is denoted

	$$ \phib_{\bf k} 
	= J_0 \left( \frac{k_\perp v_\perp}{\Omega_i} \right) \delta \phi_{\bf k}, $$
and the quasi-neutrality condition is 
\begin{equation}
    \sum_a \frac{n_a e_a^2}{T_a} \;\delta \phi_{\bf k} = e_i \int g_{\bf k} J_{0} d^3v, 
	\label{field1},
\end{equation}
where $J_{n} = J_n(k_\perp v_\perp / \Omega_i)$ is a Bessel function.  Following our previous convention, we define the free energy as twice that which appears in some other publications.  Henceforth, we suppress the $\bf k$-subscripts.

\subsection{Electrostatic energy and positive-definiteness of free energy}

It is useful to decompose the free energy into a part associated with a perturbed distribution function and a part associated with fluctuations in the electrostatic field, {\em i.e.} 

\begin{equation}
	H = G + \Ephi,\label{H-def-2}
\end{equation}
where

\begin{eqnarray}
    G = -T_i S_i = \lang  T_i \int \frac{|\delta F|^2}{F_{i0}} d^3v \rang\label{G-def-eqn}\\
    \Ephi = \lang  \left(\tau + 1 - \Gamma_{0} \right)\frac{n_i e_i^2}{T_i}|\delta \phi|^2 \rang.\label{E-def-eqn}
\end{eqnarray}
Recall the conventional definitions $\Gamma_n(b) = \exp(-b)I_n(b)$ and $b = k_\perp^2 \rho_i^2 = k_\perp^2 T_i / (m_i \Omega_i^2)$, and $\tau = (e T_i)/(e_i T_e)$.  Note that $\delta F = g - (e_i \phib/T_i) F_0$ is the gyro-averaged perturbed distribution function, and these two contributions to $H$ can be identified as the gyrokinetic perturbed entropy and the gyrokinetic field energy.

Although the general electromagnetic free energy admits a similar form as Eqn.~\ref{H-def-2} (Eqn.~3.11 Paper I), we note that the electrostatic limit is distinguished by the fact that the field contribution $\Ephi$ is itself a nonlinear invariant of the gyrokinetic system \citep{Schekochihin_2009}, by which we mean that it is conserved under the sole action of the nonlinearity, but not by all the linear terms; the same applies for the Helmholtz free energy.  The conservation of $\Ephi$ may be viewed as an additional constraint on the nonlinear dynamics, with consequences \eg for the cascade and production of large-scale $E\times B$ flows \citep{plunk-JFM-2010}.

For what follows, we need the electrostatic energy balance equation.  This is obtained by multiplying the ion gyrokinetic equation by $e_i \phib^\ast$, integrating over velocity, averaging over the parallel coordinate $l$, and summing over perpendicular wavenumber ${\bf k}$, yielding
\citep{Helander_2013}
\begin{equation}
    \frac{d}{d t}\sum_{\bf k} \Ephi = 2 \sum_{\bf k} \Dphi,\label{energy-balance-E}
\end{equation}
where the drive term $\Dphi$ is 

\begin{equation}
    \Dphi = -{\rm Re} \; e_i \lang \int \phib^\ast\left(v_{\|}\frac{\partial }{\partial l} + i\omega_{d}\right)g  d^3v \rang. 
	\label{Dphi-eqn}
\end{equation}
This is composed to two contributions, one coming from the parallel streaming term, and the other coming from the magnetic drift term.  The first contribution has a simple physical interpretation, as the rate of energy exchanged between particles and the parallel electric field ({\em i.e.} the volume average of the parallel current multiplied by the parallel electric field), while the second term describes an analogous process in the perpendicular direction associated with the drift motion of gyrocenters.

Eqn.~\ref{H-def-2} is a physically transparent form that makes it clear that the free energy $H$ is a positive-definite norm for the distribution function $g$\footnote{By the same argument, using Eqn.~\ref{field1}, $H$ can be shown to also be a positive-definite norm for the total deviation of the distribution function $\delta f = g - (e_i\delta\phi/T_i) F_0$ from the zeroth-order Maxwellian.}, {\em i.e.}

\begin{equation}
    H \geq 0,\text{ and } H = 0 \text{ iff } g = 0
\end{equation}
over all of phase space, $\ell$ and ${\bf v}$.  To see this, note that the quantities $G$ and $E$ are both positive, {\em i.e.} $G \geq 0$, obviously, and $E \geq 0$ because $\Gamma_0 \leq 1$.  Therefore if $H = 0$ then both $E = 0$ and $G = 0$.  The first implies $\delta\phi = 0$ everywhere, while the second implies $\delta F = 0$ over all of phase space; $\delta\phi = 0$ and $\delta F = 0$ obviously implies $g = 0$.

We note that positive-definiteness is a desirable property of an energetic measure that can be useful for setting bounds on the growth rate of fluctuations; if a non-zero fluctuation ($g\neq 0$) has zero measure $M$, while $dM/dt \neq 0$, then the rate $M^{-1}dM/dt$ is unbounded, so that the problem of optimal modes ({\em i.e.} determining the form of $g$ that maximizes growth) is ill-posed.  For an example of this, consider $M = E$.  One may find functions $g_\epsilon$ such that $|\delta \phi| < \epsilon$ for arbitrarily small $\epsilon > 0$, while $g_\epsilon \sim 1$ is itself not small.  In this case $dE/dt \sim \epsilon$ while $E \sim \epsilon^2$ so the rate $E^{-1}dE/dt$ is divergent in the limit $\epsilon \rightarrow 0$.

Although we mainly consider a plasma with a single kinetic ion species and adiabatic electrons, the concepts and the formalism carry over to the more general case of a plasma with an arbitrary number of kinetic species, as shown in Appendix \ref{Several-species-appx}. An important limitation, however, is that magnetic fluctuations and collisions are neglected. 

\subsection{Generalized Free Energy}\label{generalized-H-sec}

The positive definiteness of $H$ suggests a family of related quadratic energetic measures that are also positive definite.  In particular it is clear that something of the form

\begin{equation}
    \tilde{H} = H - \Delta E,\label{eq:tilde-H-form}
\end{equation}
will be positive-definite, by the same arguments of the previous section, for particular values of the parameter $\Delta$.  For instance the choice $\Delta < 1$ allows trivial generalization of the arguments, but we will see that the value can be extended beyond this.

To find a range of permissible values of $\Delta$, and to help simplify subsequent derivations, we will consider a linear transformation on the distribution function, {\em i.e.} we define a new distribution function $\tilde{g}$ in terms of which the energy $\tilde{H}$ may be expressed using the Euclidean (or $L^2$) norm,

\begin{equation}
    \tilde{H} = ||\tilde{g}||^2 =  (\tilde{g}, \tilde{g}),\label{Hg-diagonal}
\end{equation}
where we have introduced the inner product

\begin{equation}
     (\tilde{g}_1, \tilde{g}_2) =  \lang T_i \int  \frac{\tilde{g}_1^{\ast}\tilde{g}_2^{}}{F_{0}} d^3v \rang.
\end{equation}
We will refer to Eqn.~\ref{Hg-diagonal} as a ``diagonal'' form of the norm, as it does not involve additional linear operations on the distribution function, compared to the form given by Eqn.~\ref{H-def-2}.  To find the relationship between $\tilde{g}$ and $g$, we introduce the Ansatz $\tilde{g} = g - \nu J_0 F_0 e_i\delta\phi/T_i$, and substitute this into Eqn.~\ref{Hg-diagonal}.  By evaluating velocity integrals using quasi-neutrality (Eqn.~\ref{field1}) and Eqn.~\ref{E-def-eqn}, one can find the free parameter $\nu$ that yields the form Eqn.~\ref{eq:tilde-H-form}, that is

\begin{equation}
\nu = \frac{1}{\Gamma_0}\left(1+\tau - \sqrt{(1+\tau - \Gamma_0)(1+\tau -\Delta \Gamma_0)}\right),
\end{equation}
where we have taken the negative root for convenience.  Observe that in order for $\nu$ to be real, we must have

\begin{equation}
    \Delta \leq (1 + \tau)/\Gamma_0.\label{Delta-constraint}
\end{equation}
The parameter $\Delta$ can of course be negative, in which case its magnitude is unbounded.  Noting that $\Gamma_0$ generally depends on $k$, we may also assume the more restrictive $\Delta \leq (1 + \tau)$ to ensure that $\tilde{H}$ remains a nonlinear invariant.

We pause to note that the choice $\Delta = 0$ yields a novel form of the conventional (Helmholtz) free energy, immediately suggesting what can be considered as the phase-space density of free energy, namely the quantity $T_i |\tilde{g}|^2/F_0$, for which there has not yet been an expression available.\footnote{The idea for a phase-space density of free energy ({\em i.e.} a quantity that can be directly integrated over phase space to yield the total free energy) was suggested by \citet{teaca-priv}.}

It is useful now to write quasi-neutrality in terms of $\tilde{g}$,

\begin{equation}
    \frac{e_i}{T_i}\delta\phi = \frac{\lambda}{n_i}\int \tilde{g} J_{0} d^3v,\label{qn-gt-eqn}
\end{equation}
where

\begin{equation}
    \lambda = \frac{1}{\sqrt{(1+\tau -\Gamma_0)(1+\tau -\Delta\Gamma_0)}}.\label{alpha-eqn}
\end{equation}

Finally, we can show that $\tilde{H}$ is positive-definite.  First, positivity follows from Eqn.~\ref{Hg-diagonal}, and it is obvious from Eqn.~\ref{field1} that if $g = 0$ then $\delta \phi = 0$ so that $E$ and $H$ both vanish, implying $\tilde{H} = 0$.  On the other hand, if we assume that $\tilde{H} = 0$, then Eqn.~\ref{Hg-diagonal} implies that $\tilde{g} = 0$, and Eqn.~\ref{qn-gt-eqn} implies that $\delta\phi = 0$, from which we conclude $g = 0$.  In summary, $\tilde{H} \geq 0$ and $\tilde{H} = 0$ iff $g = 0$.

\section{Modes of optimal growth}

A key point in introducing the generalization of free energy $\tilde{H}$ is that this quantity introduces wave-particle effects (parallel resonance and drift resonance) that enter the electrostatic energy balance equation, Eqn.~\ref{energy-balance-E}.  

We note that for the choice $\Delta = 0$, the energy $\tilde{H}$ reduces to the conventional Helmholtz free energy that we have studied in the previous papers.  For this choice the modes of optimal growth correspond exactly to the modes introduced in section 6 of Part I, and studied in fully electromagnetic limit in \citet{plunk_helander_2022} (Part II).  Because those modes are included as a limit of our present analysis, we can be assured that the most stringent bound on growth obtained from the generalized free energy will be at least as good as the known bound obtained from the Helmholtz free energy.

Note that, as long as the parameter $\Delta$ is  independent of ${\bf k}$, the quantity $\tilde{H}$ is conserved by the nonlinearity, {\em i.e.} under summation over ${\bf k}$.  This is because it is a linear combination of two nonlinear invariants.  One simply combines Eqns.~\ref{energy-balance-H} and \ref{energy-balance-E} to obtain

\begin{equation}
    \frac{d}{d t}\sum_{\bf k} \tilde{H} = 2 \sum_{\bf k} (D - \Delta \Dphi),\quad\text{for $\Delta$ independent of ${\bf k}$,}\label{energy-balance-Hg}
\end{equation}
{\em i.e.} the change of this measure is due to the drive terms of electrostatic and free energy, and is otherwise conserved by the turbulent interactions.  It is potentially useful to also consider $\Delta$ that does depend on ${\bf k}$, for the purpose of obtaining bounds on linear growth, but the nonlinear implications will be less clear in that case.

In direct analogy to how modes of optimal free energy growth were defined, we introduce a rate $\Lambda$

\begin{equation}
    \Lambda = (D-\Delta\Dphi)/\tilde{H}\label{Lambda-eqn}
\end{equation}
 to be optimized over the space of ion distribution functions $g$.  We note the bound on conventional gyrokinetic instability growth rates,

\begin{equation}
    \gamma_L \leq \max_{g} \Lambda.\label{eq:linear-bound}
\end{equation}

Having already found a diagonal form of the generalized free energy, Eqn.~\ref{Hg-diagonal}, we need not use a variational approach to find the states of extremal $\Lambda$.  We simply identify the Hermitian linear operators associated with the input of free energy and electrostatic energy, {\em i.e.}

\begin{eqnarray}
    D = (\tilde{g}, {\cal D} \tilde{g}),\\
    \Dphi = (\tilde{g}, {\cal K} \tilde{g})
\end{eqnarray}
To obtain these forms note that, from Eqn.~\ref{qn-gt-eqn}, $\delta \phi$ can be regarded as the result of a linear operator on $\tilde{g}$.  Then, using Eqn.~\ref{qn-gt-eqn} and Eqns.~\ref{D-eqn} and \ref{Dphi-eqn}, and some straightforward algebra (see Appendix \ref{operator-derivations-appx}), we obtain explicit forms for the operators.  First we have

\begin{equation}
    {\cal D} \tilde{g} = \frac{i\lambda}{2 n_i} J_0 F_0 \osT \int d^3v^\prime J_0^\prime \tilde{g}^\prime\left[ \left(\frac{v}{\vth}\right)^2 - \left(\frac{v^\prime}{\vth}\right)^2\right],\label{D-op-eqn}
\end{equation}
where primes denote evaluation at ${\bf v}^\prime$ and $\vth = \sqrt{2 T_i/m_i}$.  For convenience, the operator ${\cal K}$ can be split into its parallel and perpendicular components as ${\cal K} = {\cal K}_{\|} + {\cal K}_d$, for which we obtain

\begin{equation}
    {\cal K}_d \tilde{g} = \frac{i\lambda}{2 n_i}\omega_d(\ell) F_0 J_0 \int d^3v^\prime J_0^\prime \tilde{g}^\prime \left[\left(\frac{\vperp}{\sqrt{2}\vth}\right)^2 + \left(\frac{\vpar}{\vth}\right)^2 - \left(\frac{\vperp^\prime}{\sqrt{2}\vth}\right)^2 - \left(\frac{\vpar^\prime}{\vth}\right)^2\right] ,\label{K-op-perp-eqn}
\end{equation}
and

\begin{multline}
    {\cal K}_{\|} \tilde{g} = \frac{\lambda}{2 n_i}F_0\left\{ J_0\left[-B\frac{\partial}{\partial l}\left(\frac{1}{B}\int d^3v^\prime \vpar^\prime J_0^\prime \tilde{g}^\prime\right) + \int d^3v^\prime \vpar^\prime \frac{\partial J_0^\prime}{\partial l} \tilde{g}^\prime \right]\right.\\
    \left.+ \frac{\vpar}{\lambda}\frac{\partial}{\partial l}\left(J_0 \lambda\int d^3v^\prime J_0^\prime\tilde{g}^\prime \right) \right \}.\label{K-op-par-eqn}
\end{multline}
In deriving Eqn.~\ref{K-op-par-eqn}, it is important to note that the parallel derivative is taken at fixed magnetic moment and particle energy, and that the velocity-space volume element $d^3v$ is proportional to $B/v_\|$ in these variables. More details are given in Appendix \ref{Several-species-appx}. The kinetic eigenvalue problem can be stated now as

\begin{equation}
    \Lambda \tilde{g} = \left({\cal D} - \Delta {\cal K}\right)\tilde{g},\label{kinetic-eigenproblem}
\end{equation}
where solutions, {\em i.e.} pairs $\Lambda_n$ and $\tilde{g}_n(l, {\bf v})$, realize optimal growth of $\tilde{H}$.  To see this, we can decompose the distribution function $\tilde{g}$ in terms of the orthogonal eigenmodes of Eqn.~\ref{kinetic-eigenproblem}, $\tilde{g} = \sum_n c_n \tilde{g}_n$ to obtain (choosing $||\tilde{g}_n|| = 1$)

\begin{equation}
    \frac{1}{2\tilde{H}}\frac{d \tilde{H}}{dt} = \frac{\sum_n |c_n|^2 \Lambda_n}{\sum_n |c_n|^2}\nonumber
\end{equation}
from which it is clear that the rate of energy growth of the system is maximized by setting the distribution function equal to the mode of largest growth rate, {\em i.e.}

\begin{equation}
    \max_{g} \frac{1}{2\tilde{H}}\frac{d \tilde{H}}{dt} = \max_{n} \Lambda_n = \Lambda_\text{max},\nonumber
\end{equation}
which bounds the normal growth of the system, $\gamma_L$, in accordance with Eqn.~\ref{eq:linear-bound}.
\subsection{Moment form of eigenproblem}

The analysis of Eqn.~\ref{kinetic-eigenproblem} is greatly simplified by adopting a moment form. 
 As found in the preceding papers, there are natural moments that appear in the energy input terms that can be identified to reduce the dimensionality of the problem substantially.  Upon inspecting the energy balance equations one finds the following key dimensionless integrals:

\begin{eqnarray}\label{moments-def}
    \kappa_1 &= &\int d^3 v J_0 \tilde{g}/n_i,\\
    \kappa_2 &= &\int d^3 v \left(\frac{v^2}{\vth^2}\right) J_0 \tilde{g}/n_i,\\
    \kappa_3 &= &\int d^3 v \left(\frac{\vperp^2}{2\vth^2} + \frac{\vpar^2}{\vth^2}\right) J_0 \tilde{g}/n_i,\\
    \kappa_4 &= &\int d^3 v \left(\frac{\vpar}{\vth}\right)J_0 \tilde{g}/n_i,\\
    \kappa_5 &= &\int d^3 v \left(\frac{\vpar}{\vth}\right) \frac{\partial J_0}{\partial l} \tilde{g}/n_i,
\end{eqnarray}
where $\kappa_1$ is a density-like moment, $\kappa_2$ and $\kappa_3$ are pressure-like, $\kappa_4$ is parallel ion flow, while $\kappa_5$ is more abstract.  

It is easy to recognize these integrals on the right hand side of Eqns.~\ref{D-op-eqn}, \ref{K-op-perp-eqn} and \ref{K-op-par-eqn}, and straightforward to rewrite those equations in moment form.  The dimensional reduction is achieved by taking moments of the  these equations to obtain a coupled set of five fluid equations.  These, which are given in Appendix \ref{moment-forms-appx}, can be combined, leading, after lengthy algebra, to a relatively simple second order ordinary differential equation, the main result of this paper:

\begin{multline}
    \left(\frac{4\Lambda^2}{\lambda^2} + \left(\Delta \omega_d G_{3}  - \osT G_{1} \right)^2 - G_{0}\left[ (\osT)^2 G_{2} - 2 \Delta \omega_d \osT G_{4} + \Delta^2\od^2 G_{5}\right] \right) \varphi \\
    = \Delta^2 \vth^2 G_{0} B \left[ -\frac{\partial}{\partial l}\left(\frac{ G_{0,2}}{B}\frac{\partial \varphi}{\partial l}\right) + \varphi\frac{ G_{0,2}^{\prime\prime}}{B} - \varphi\frac{\partial}{\partial l}\left(\frac{ G_{0,2}^{\prime}}{B}\right) \right ],\label{kappa1-ode-eqn}
\end{multline}
where $\varphi = e_i \delta\phi/T_i = \lambda \kappa_1$ is the normalized electrostatic potential. 
 The functions $G_{m,n}$, $G^\prime_{m,n}$, and $G^{\prime\prime}_{m,n}$, which depend on arc length via $b(l)$ and $B(l)$, are defined in terms of integrals involving Bessel functions, and are evaluated in Appendix \ref{bessel-integrals-appx}; see in particular Eqns.~\ref{eq:Gperp}-\ref{eq:Gmn} and \ref{eq:GperpPrime}-\ref{eq:GPrimePrime}.  The other $b$-dependent factors ($G_0$-$G_5$) can be expressed in terms of $G_{m,n}$, and are evaluated in terms of more elementary Bessel functions in Appendix \ref{bessel-integrals-appx-b}.

In Eqn.~\ref{kappa1-ode-eqn} we see the eigenvalue $\Lambda$ entering quadratically, reflecting the fact that there will be two real roots, one positive and one negative, owing to Hermiticity and time-reversal symmetriy of the full eigenproblem, Eqn.~\ref{kinetic-eigenproblem}.  Note that the terms arising from the parallel drive of electrostatic energy are placed on the right hand side.  In the following section, we will consider some simple limits of this equation, and leave its more general solution for a future publication.

\section{Simple limits}

In this section we will consider some simple limits applied to Eqn.~\ref{kappa1-ode-eqn}, and draw some comparison to linear theory of the main instability targeted by limit of this paper, the ion temperature gradient (ITG) mode (see for instance \citet{plunk-POP-2014}).  To start, we note that taking $\Delta = 0$, so that $\tilde{H}$ becomes the conventional Helmholtz free energy, yields

\begin{equation}
\Lambda^2 = \frac{(\osT)^2}{4(1+\tau -\Gamma_0)(1+\tau)} \left( G_{0}G_{2} - G_{1} ^2\right),
\end{equation}
which matches Eqn.~6.20 of Paper I.

In considering other simplifications, we first should note that the adiabatic electron approximation already neglects the contribution from a trapped electron population, which requires either large electron collisionality, or uniform magnetic field strength as measured along the field line.  Let us assume the latter for simplicity,

\begin{equation}
    \frac{\partial B}{\partial l} = 0.
\end{equation}
Making this assumption simplifies somewhat Eqn.~\ref{kappa1-ode-eqn}, where all the explicit factors of $B$ drop out of the right-hand side.  A more significant simplification is achieved by assuming unsheared and uniform magnetic geometry, in particular

\begin{eqnarray}
    \frac{\partial b}{\partial l} = 0,\\
    \frac{\partial \od}{\partial l} = 0,
\end{eqnarray}
In this limit, all of the coefficients of Eqn.~\ref{kappa1-ode-eqn} are constants, and a simple dispersion relation is the obtained by taking $\partial \varphi/\partial l = i k_\| \varphi$.  We find

\begin{equation}
    \frac{4\Lambda^2}{\lambda^2} + \left(\Delta \omega_d G_{3}  - \osT G_{1} \right)^2 - G_{0}\left[ (\osT)^2 G_{2} - 2 \Delta \omega_d \osT G_{4} + \Delta^2\od^2 G_{5}\right] = \Delta^2 k_\|^2 \vth^2 G_{0}^2/2.\label{full-disp}
\end{equation}
were we have used $G_{0,2} = G_{0}/2$.  As noted in Section \ref{generalized-H-sec}, the quantity $\Delta$ is a free parameter, over which we can optimize $\Lambda$ to improve the bounds on the growth rate of fluctuations.

\subsection{Slab ITG mode}\label{slab-itg-sec}

Setting $\od = 0$ leaves only the slab branch of the ITG mode, driven by the temperature gradient, and involving ion parallel resonance.  Eqn.~\ref{full-disp} reduces to

\begin{equation}
    \frac{4\Lambda^2}{(\osT)^2\lambda^2} = G_{0}G_{2} - G_{1}^2 + \Delta^2 \kappa_\|^{-2} G_{0}^2/2,\label{slab-itg-disp}
\end{equation}
where $\kappa_\| = \osT/(k_\|\vth)$.  Because $G_0G_2 - G_1^2 \geq 0$, the two contributions on the right hand side are both positive but the solution for which $\Lambda$ is minimal is actually not obtained for $\Delta = 0$, due to the implicit dependence of $\lambda$ on $\Delta$ given by Eqn.~\ref{alpha-eqn}.

Although all values of the parameter $\Delta$ satisfying Eqn.~\ref{Delta-constraint} yield a valid bound on the growth rate of normal modes ($\gamma_L \leq \Lambda$), the lowest value is the most stringent and serves as the closest approximation of $\gamma_L$.  To obtain this ``optimal bound'', we can consider the extrema of $\Lambda^2/(\osT)^2$, {\em i.e.}

\begin{equation}
    \frac{d}{d\Delta}\left( \frac{G_{0}G_{2} - G_{1}^2 + \Delta^2 \kappa_\|^{-2} G_{0}^2/2}{(1+\tau -\Gamma_0)(1+\tau -\Delta\Gamma_0)}\right) = 0.\label{Delta-extreme-eqn}
\end{equation}
This results in a quadratic equation for $\Delta$ that is still rather complicated so we will consider the limit $b \rightarrow 0$; see Appendix \ref{small-b-limit-appx} for the relevant limits of $G_{m,n}$, {\em etc}. Applying the limit to Eqn.~\ref{slab-itg-disp} yields

\begin{equation}
    \frac{\Lambda^2}{(\osT)^2} = \frac{3 + \Delta^2/\kappa_\|^2}{8\tau(1 + \tau - \Delta)}\label{slab-itg-small-b}
\end{equation}

This solution diverges as $\Delta$ approaches $1+\tau$; recall that this is the upper limit allowed by  Eqn.~\ref{Delta-constraint}.  It also grows in an unbounded fashion as $\Delta \rightarrow -\infty$.  There is an optimal value giving minimal $|\Lambda|$, obtained by solving Eqn.~\ref{Delta-extreme-eqn} in this limit.  This solution, denoted as $\Delta_\text{min}$, is

\begin{equation}
    \Delta_\text{min} = 1 + \tau - \sqrt{(1+\tau)^2 + 3 \kappa_\|^2}
\end{equation}
where the negative root has been selected to be consistent with Eqn.~\ref{Delta-constraint}.  Substituting this solution into Eqn.~\ref{slab-itg-small-b} gives

\begin{equation}
    \Lambda_{\min}^2 = \frac{(\osT)^2}{4\bar{\kappa}_\|^2 \tau(1+\tau)}\left( \sqrt{1 + 3 \bar{\kappa}_\|^2} - 1\right),\label{eq:slab-optimal-rate}
\end{equation}
where we define $\bar{\kappa}_\| = \kappa_\|/(1+\tau)$.  This reaches its maximum value in the limit $\bar{\kappa}_\| \rightarrow 0$, and is a decreasing function of $|\bar{\kappa}_\||$, {\em i.e.}

\begin{equation}
    \Lambda_{\min}^2 = \begin{cases}
			\frac{3}{8 \tau  (\tau +1)}(\osT)^2, & \text{for $\bar{\kappa}_\| \rightarrow 0$,}\\
            \frac{\sqrt{3}}{4\tau}|\osT k_\parallel \vth|, &\text{for $|\bar{\kappa}_\|| \gg 1$},
		 \end{cases}
\end{equation}
Physically, the first result implies that when drive ($\osT$) is much smaller than the parallel transit frequency ($k_\parallel\vth$), the best bound is equal to that obtained by free energy ($\Delta = 0$).  In this case, the bound is consistent from expectations of the growth rate of a resonant slab ITG mode, {\em i.e.} $\gamma_L \sim \osT$.

In the opposite limit ($\bar{\kappa}_\| \gg 1$), however, when the drive large, {\em i.e.} in the so-called non-resonant or ``fluid'' limit, we obtain a much lower bound, essentially the geometric mean of the drive and the parallel transit frequency $k_\|\vth$.  We note that this bound is not as low as what is obtained from the non-resonant solution of the dispersion relation (without density gradient), {\em i.e.} $\gamma_L \sim \osT^{1/3}(k_\parallel\vth)^{2/3}$ \citep{plunk-POP-2014}, but nevertheless captures the expected weakening (relative to the resonant result) qualitatively.  A plot of the optimal bound of Eqn.~\ref{eq:slab-optimal-rate} is provided in Figure \ref{fig:tor-growth-bound}.  Note that the resonant stabilization at high $k_\|$ (low $\bar{\kappa}_\|$) is not captured in this case.

\begin{figure}
    \centering
    \includegraphics{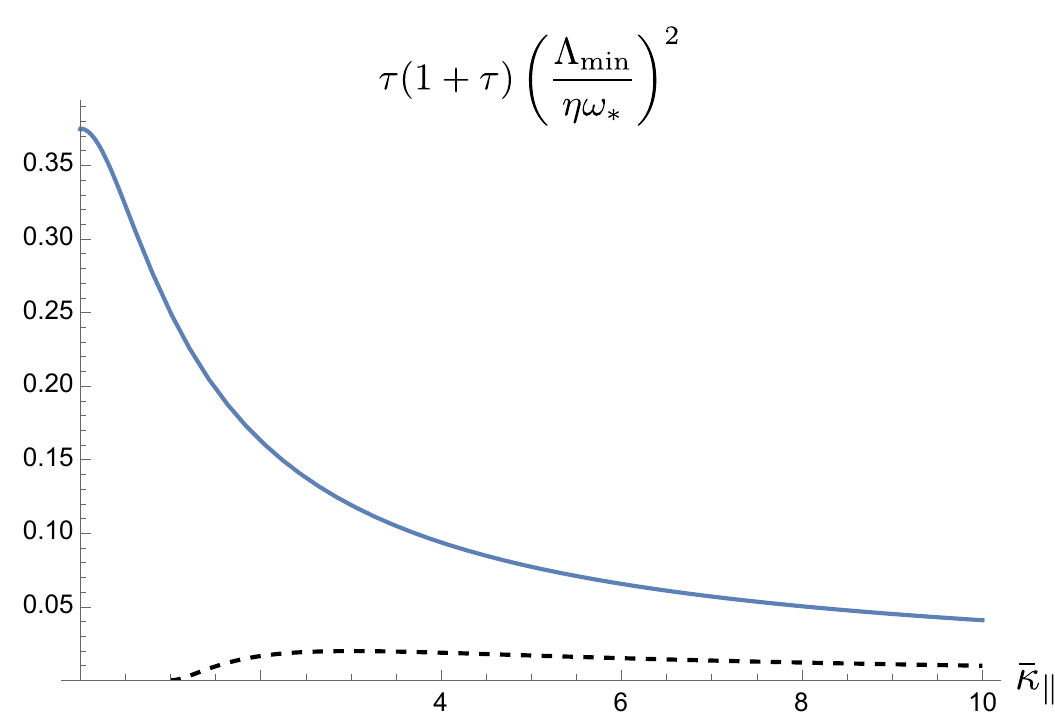}
    \caption{Bound of the growth rate of the slab ITG mode (blue), obtained from the optimal growth of generalized free energy, plotted versus the instability parameter $\bar{\kappa}_\| = \osT/[k_\parallel\vth(1+\tau)]$.  For comparison the growth rate $\gamma_L$ is obtained by solving the linear dispersion relation (Eqn.~\ref{eq:local-disp-reln}), and the quantity $\tau (1+\tau) (\gamma_L/\eta\omega_*)^2$ is plotted for the case $\tau = 1$ (dashed black).}
    \label{fig:slab-growth-bound}
\end{figure}

It is interesting to observe that this latter limit corresponds to $\Delta_\text{min} \rightarrow  -\infty$, making $\tilde{H}$ in some sense dominated by the electrostatic component.

\subsection{Toroidal ITG mode}

Now taking $k_\parallel\vth$ to be small, we can neglect the right-hand side of Eqn.~\ref{full-disp}, leaving
\begin{equation}
    \frac{4\Lambda^2}{\lambda^2} = G_{0}\left[ (\osT)^2 G_{2} - 2 \Delta \omega_d \osT G_{4} + \Delta^2\od^2 G_{5}\right] - \left(\Delta \omega_d G_{3}  - \osT G_{1} \right)^2.\label{toroidal-disp}
\end{equation}
To derive the optimal choice of $\Delta$, we again take the $b \rightarrow 0$ limit and obtain from Eqn.~\ref{toroidal-disp}

\begin{equation}
    \frac{\Lambda^2}{(\osT)^2} = \frac{3 \Delta ^2-8 \Delta  \kappa_d +6 \kappa_d^2}{16 \kappa_d^2 \tau  (\tau + 1 - \Delta)},\label{tor-disp-small-b}
\end{equation}
where we define $\kappa_d = \osT/\od$.  Note the similar qualitative behavior with $\Delta$ as Eqn.~\ref{slab-itg-disp}, namely its divergence at the $\Delta \rightarrow 1 + \tau$, and unbounded growth as $\Delta \rightarrow -\infty$.  The key difference here arises in the linear term in the drive parameter $\kappa$; this is expected from the theory of the toroidal ITG mode since the sign of the drift frequency (associated with so-called `good' and 'bad' magnetic curvature) is important for the resonance.

\begin{figure}
    \centering
    \includegraphics{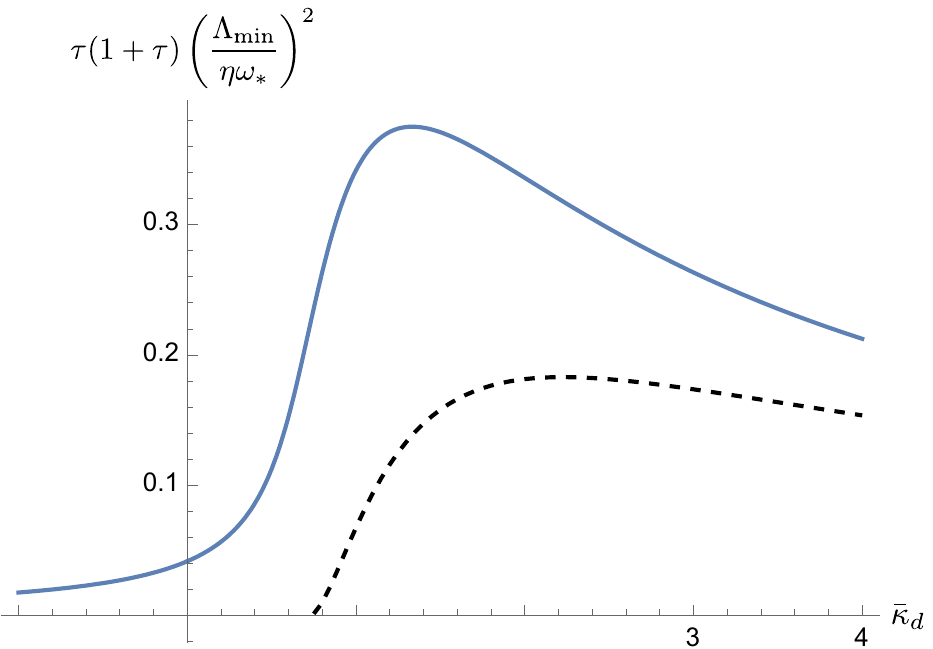}
    \caption{Bound of the growth rate of the toroidal ITG mode, obtained from the optimal growth of generalized free energy, plotted versus the instability parameter $\bar{\kappa}_d = \osT/[\od(1+\tau)]$.  For comparison the growth rate $\gamma_L$ is obtained by solving the linear dispersion relation (Eqn.~\ref{eq:local-disp-reln}), and the quantity $\tau (1+\tau) (\gamma_L/\eta\omega_*)^2$ is plotted for the case $\tau = 1$ (dashed black).}
    \label{fig:tor-growth-bound}
\end{figure}

We now find the value of $\Delta$ that minimizes $\Lambda$:

\begin{equation}
\Delta_\text{min} = (1+\tau) \left(1 - \sqrt{2 \bar{\kappa}_d^2-8\bar{\kappa}_d/3 +1}\right),
\end{equation}
where we define the parameter $\bar{\kappa}_d = \kappa_d/(1 + \tau)$.  Substituting this into Eqn.~\ref{tor-disp-small-b} yields

\begin{equation}
    \Lambda_\text{min}^2 = (\osT)^2\left(  \frac{8 \bar{\kappa}_d \left(\zeta \left(\bar{\kappa}_d\right)-1\right)+3 \left(\zeta
   \left(\bar{\kappa}_d\right)-1\right)^2+6 \bar{\kappa}_d^2}{16 \tau  (\tau +1)
   \bar{\kappa}_d^2 \zeta \left(\bar{\kappa}_d\right)} \right),\label{tor-disp-small-b-opt}
\end{equation}
with $\zeta = \sqrt{2 \bar{\kappa}_d^2-8\bar{\kappa}_d/3 + 1}$.  This expression for $\Lambda_\text{min}$ is naturally separated into a factor that depends only on the instability parameter $\bar{\kappa}_d$, from which we can derive the asymptotic behavior.  To show the behavior of this factor we plot the quantity $\tau(1 + \tau)\Lambda_\text{min}^2/(\osT)^2$ in Fig.~\ref{fig:tor-growth-bound}.  The overall behavior of  $\Lambda_\text{min}$ is captured by the following limits

\begin{equation}
    \Lambda_\text{min}^2 = \begin{cases}
			|\osT\od|\left(\frac{3 \sqrt{2} + 4 \sigma}{8 \tau}\right), & \text{for $|\bar{\kappa}_d| \gg 1$,}\\
            \frac{(\osT)^2}{24\tau(1+\tau)}, &\text{for $\bar{\kappa}_d \rightarrow 0$},\\
             \frac{3(\osT)^2}{8\tau(1+\tau)}, &\text{for $\bar{\kappa}_d \rightarrow 4/3$}
		 \end{cases}
\end{equation}
where we denote $\sigma = \pm 1$ as the sign of $\kappa_d$.  At large drive ($|\bar{\kappa}_d| \gg 1$; $|\osT| \gg |\od| (1+\tau)$) we recover the expected non-resonant (``fluid'') behavior of the toroidal ITG mode with no density gradient, namely $\gamma_L \sim \sqrt{\osT\od}$. Note that this growth rate is much smaller than the bound found by merely considering the Helmholtz free energy \citep{helander_plunk_2022}. Although we do not see the complete stabilization ($\Lambda = 0$) at negative values of $\bar{\kappa}_d$ (opposite sign of $\od$ and $\osT$) expected from theory, there is a strong asymmetry, with $|\Lambda|$ having its larger values at positive $\bar{\kappa}_d$ and being comparatively much smaller for negative $\bar{\kappa}_d$.

The value $\bar{\kappa}_d = 4/3$ ({\em i.e.} $\osT = 4 (1+\tau)\od/3$) achieves the maximal value of $\Lambda_\text{min}$ at fixed $\osT$, and therefore is evocative of the resonance condition for the toroidal ITG modes $\osT \sim \od$ \citep{biglari}.  This value of $\bar{\kappa}_d$ is obtained by solving for $\Delta_\text{min} = 0$, explaining why it produces the worst bound, {\em i.e.} that given by optimal growth of Helmoltz free energy.

It is noteworthy that for the limit $\bar{\kappa}_d \rightarrow 0$ ($\omega_d \gg \osT/(1+\tau)$) our method yields a value of $|\Lambda|$ that is a factor of $1/3$ reduced as compared to the resonant case, again at least qualitatively reproducing the expected stabilization of the toroidal ITG mode in this limit.

\section{Conclusion}

We have demonstrated that the use of a generalized form of free energy $\tilde{H}$ introduces some of the physics of wave-particle resonance that is missing in the theory of optimal mode growth of Helmholtz free energy \citep{helander-plunk-prl-2021, helander_plunk_2022, plunk_helander_2022}. The growth rates of optimal modes of generalized free energy provide a rigorous upper bound on the growth of conventional gyrokinetic instabilities (``normal modes''), which is always below the Helmholtz bound, as it must be, given that the Helmholtz free energy is a special case of the generalized measure. Moreover, optimal modes of generalized free energy depend on the magnetic-field geometry to a greater extent than those associated with Helmholtz free energy. The difference in growth rates can be very large. For instance, in the important case of a strongly driven toroidal ITG mode, the Helmholtz bound is larger by a factor of order $\eta \omega_\ast / \omega_d \gg 1$. 

A single ordinary differential equation has been derived for optimal modes, allowing general magnetic geometry.  We found solutions of this equation in some simple limits to demonstrate that it indeed recovers, at least qualitatively, some of the physical effects expected from the theory of linear ITG modes, including sensitivity to the ratio of the frequencies associated with drive and resonance, and transition of the instability when this ratio is near one.  Density gradient dependence of ITG mode is absent from both electrostatic and free energy input terms, assuming adiabatic electrons, so its effect is not accounted for by the theory presented here.

The results of this work have possible implications for ``turbulence optimization'', {\em i.e.} the endeavor to shape the equilibrium magnetic geometry of stellarators for low turbulence. The general result, Eqn.~\ref{kappa1-ode-eqn}, allows in principle for the inclusion of the complete geometric information that is needed to run gyrokinetic simulations.  However, the solution of this equation should be far simpler and more efficient, due to the reduction of velocity space to a single moment.  Stellarator optimization requires a large number of calculations to be performed, which so far has severely limited the use of direct gyrokinetic simulations, even linear ones.  The simplifications obtained here, which also include the removal of the need for time integration, will need to be balanced against numerical costs associated with eigenvalue problems, and possible costs associated with determining the optimal value of the parameter $\Delta$, if the absolute strongest bound is desired.

Our results found for the toroidal branch of the ITG mode hint at a possible optimization strategy based on optimal modes.  Consider fixed plasma conditions, {\em i.e.} a given temperature gradient ($\osT$) and temperature ratio ($\tau$):  At high drive ($\osT/(1 + \tau) > 4\od/3$), minimization of the optimal growth rate $|\Lambda|$ is achieved by minimization of the magnetic drift $\od$ ({\em i.e.} magnetic curvature), corresponding to minimization of the strongly-driven (non-resonant) toroidal ITG mode.  On the other hand, at low drive ($\osT/(1 + \tau) < 4\od/3$) the increase of $\od$ is favored, corresponding to a weakening of the marginally unstable ITG mode, {\em i.e.} an increase of the threshold of instability.  The latter case corresponds to ``critical gradient'' optimization, an idea which has recently been developed \citep{Roberg-Clark-PRR-2022, Roberg-Clark-HSK-2022}.

It is worth mentioning the larger context in which optimal modes are potentially interesting to study, and applications besides their use to bound or estimate the growth rate of normal instabilities.  Although these modes ($\tilde{g}_n$, $\Lambda_n$) do not arise as late-time-asymptotic solutions of initial-value gyrokinetic simulations, as normal modes do, they are still realizable in the sense that a gyrokinetic simulation initialized in one of these states $\tilde{g}_n$ will temporarily exhibit energy growth rate exactly equal to $2\Lambda_n$.  These modes modes are thus just as ``real'' as conventional eigenmodes and may also be observed, to some extent, in nonlinear simulations, where fluctuations are continually driven away by nonlinear interactions from the form of normal linear instabilities.  They are also potentially useful for studying systems without any unstable normal modes, {\em i.e.} ``subcritical'' turbulence (indeed this is the context in which they were originally formulated).  Furthermore, since optimal modes are based on quadratic norms that are nonlinearly conserved by the gyrokinetic equations, they can be used to bound the instantaneous growth rates observed in fully nonlinear solutions; this point was made in Parts I and II, but applies equally well here.

More general solutions of the optimal mode equation, and the application to optimization will be pursued in future works.  Other special limits can also be explored including the limit of large electron-bounce frequency, appropriate for studying trapped-electron modes, or adiabatic-ion limits applied to universal instabilities.  Electromagnetic generalizations are also possible: although it is not clear how to construct a positive-definite electromagnetic form of generalized free energy $\tilde{H}$ that is a nonlinear invariant, it is certainly possible to consider related measures that focus on linear bounds.

{\bf Funding.}  This work has been carried out within the framework of the EUROfusion Consortium, funded by the European Union via the Euratom Research and Training Programme (Grant Agreement No 101052200 — EUROfusion). Views and opinions expressed are however those of the author(s) only and do not necessarily reflect those of the European Union or the European Commission. Neither the European Union nor the European Commission can be held responsible for them.  This work was partly supported by a grant from the Simons Foundation (560651, PH).

{\bf Declaration of Interests}. The authors report no conflict of interest.

{\bf Author ORCID.} G. G. Plunk, https://orcid.org/0000-0002-4012-4038; P. Helander, https://orcid.org/0000-0002-0460-590X.

\appendix

\section{Several kinetic species}\label{Several-species-appx}

For a plasma with an arbitrary number of particle species, we multiply each gyrokinetic equation
\begin{eqnarray}
  \frac{\partial g_{a,{\bf k}}}{\partial t} +
	v_{\|} \frac{\partial g_{a,{\bf k}}}{\partial l} + i \omega_{da} g_{a, {\bf k}} 
	+ \frac{1}{B^2} \sum_{{\bf k}'} {\bf B} \cdot ({\bf k} \times {\bf k}')  
	\phib_{{\bf k}'} g_{a, {\bf k} - {\bf k}'} \\
 =  \frac{e_a F_{a0}}{T_a} 
	\left( \frac{\partial}{\partial t} + i \omega_{\ast a}^T \right) \phib_{\bf k}\; 
	\label{gk},
	\end{eqnarray}
by $e_a \phib_{\bf k}^\ast$, integrate over velocity space, take the real part and the average $\langle \cdots  \rangle$ over the flux tube, and sum over all species $a$ and wave vectors $\bf k$. In other words, we apply the operator
    \begin{eqnarray} {\rm Re} \; \sum_{a,{\bf k}} e_a \left\langle \int  
    \phib_{\bf k}^\ast \left(  \cdots \right) d^3v \right\rangle. 
    \end{eqnarray}
Since the expression
    \begin{eqnarray}
    {\rm Re} \; ({\bf k} \times {\bf k}') \phib_{{\bf k}'}^\ast \phib_{\bf k} 
    g_{a,{\bf k} - {\bf k}'}
    = \frac{1}{2} ({\bf k} \times {\bf k}') 
    \left( \phib_{{\bf k}'}^\ast \phib_{\bf k} 
    g_{a,{\bf k} - {\bf k}'} + \phib_{{\bf k}'} \phib_{\bf k}^\ast 
    g_{a,{\bf k} - {\bf k}'}^\ast \right) 
	\end{eqnarray}
    \begin{eqnarray}
    = \frac{1}{2} ({\bf k} \times {\bf k}') 
    \left( \phib_{-{\bf k}'} \phib_{\bf k} 
    g_{a,{\bf k} - {\bf k}'} + \phib_{{\bf k}'} \phib_{-\bf k} 
    g_{a,{-\bf k} + {\bf k}'} \right) 
	\end{eqnarray}
changes sign under an exchange of  $\bf k$ and ${\bf k}'$, the nonlinear terms cancel upon summation over $\bf k$ and ${\bf k}'$, and we obtain
    \begin{eqnarray}
  {\rm Re} \; \sum_{a,{\bf k}} e_a \left\langle  \int  
    \phib_{\bf k}^\ast \left(
  \frac{\partial g_{a,{\bf k}}}{\partial t} +
	v_{\|} \frac{\partial g_{a,{\bf k}}}{\partial l} + i \omega_{da} g_{a, {\bf k}} 
 -  \frac{e_a F_{a0}}{T_a} 
	\frac{\partial \phib_{\bf k}}{\partial t} \right) d^3v 
	\right\rangle = 0.
	\end{eqnarray}
The quasineutrality equation (\ref{field1}) can be used to write the first term as
    \begin{eqnarray}
  {\rm Re} \; \sum_{a,{\bf k}} e_a \int  
    \phib_{\bf k}^\ast   \frac{\partial g_{a,{\bf k}}}{\partial t} d^3v
    = \frac{d}{dt} \sum_{\bf k} 
    \frac{n_a e_a^2}{2 T_a} \left| \delta \phi_{\bf k} \right|^2.
    \end{eqnarray}
We thus arrive at the electrostatic energy balance equation
        \begin{eqnarray}
    \frac{d}{dt} \sum_{\bf k} 
    \frac{n_a e_a^2}{2 T_a} \left\langle \left[1 - \Gamma_0(b_{a{\bf k}}) \right]
    \left| \delta \phi_{\bf k} \right|^2
    \right\rangle 
    \\= - {\rm Re} \; \sum_{a,\bf k} 
    \left\langle e_a \int \phib_{\bf k}^\ast 
    \left( v_\| \frac{\partial g_{a,{\bf k}}}{\p l} + i \omega_{da} g_{a,{\bf k}} \right) d^3v \right\rangle,
        \end{eqnarray}
which is the generalization of Eqn.~\ref{energy-balance-E} to several species. The right-hand side can be interpreted as minus the work done by the electric field on the various particle species. The first term in this expression contains
    \begin{eqnarray}  
    \int J_{0a} v_\| \frac{\partial g_{a,{\bf k}}}{\p l} d^3v
    = \sum_\sigma \frac{2 \pi \sigma B}{m_a^2} \int_0^\infty dE_a
    \int_0^{E_a/B} J_{0a} \frac{\partial g_{a,{\bf k}}}{\p l} d\mu_a
    \\ = B \frac{\partial}{\partial l} \left( \frac{1}{B}
    \int J_{0a} v_\| g_{a,{\bf k}} d^3v \right)
    - \int \frac{\partial J_{0a}}{\partial l} v_\| g_{a,{\bf k}} d^3v, 
    \end{eqnarray}
which we used in Eqn.~\ref{K-op-par-eqn}, with $\sigma = v_\| / | v_\| |$, $E_a = m_av^2/2$ and $\mu_a = m_a v_\perp^2/(2B)$. 

\section{Derivation of operators ${\cal D}$ and ${\cal K}$}\label{operator-derivations-appx}
We first write forms of $D$ and $\Dphi$ explicit in $\tilde{g}$, noting that the contribution to $D$ proportional to the density gradient is zero by use of quasi-neutrality with the adiabatic electron approximation (the factor $\osT \propto dT_i/d\psi$ appears in what follows, but never $\omega_\ast \propto dn_i/d\psi$ alone).  The terms proportional to $\nu$, involved in the transformation from $g$ to $\tilde{g}$, are also zero, due to oddness in $v_\|$, so expressing energy input in terms of $\tilde{g}$ merely has the consequence of introducing the overall factor $\lambda$.  The expressions are

\begin{eqnarray}
    D = - \lang T_i\frac{i\lambda }{2 n_i} \; \int \tilde{g}({\bf v})\tilde{g}^*({\bf v}^\prime)\;\osT  \left(\frac{v^2}{\vth^2}\right) J_0 J_0^\prime d^3v d^3v^{\prime}\rang + \text{c.c.},\\
    K_\| = - \lang T_i\frac{\lambda}{2n_i} \; \int \tilde{g}^*({\bf v}^\prime)\left(v_{\|}\frac{\partial \tilde{g}({\bf v})}{\partial l}\right) J_0 J_0^\prime d^3v d^3v^\prime \rang + \text{c.c.},\\
    K_d = - \lang T_i\frac{i\lambda}{2n_i} \; \int \tilde{g}^*({\bf v}^\prime)\omega_{d}\tilde{g}({\bf v}) J_0 J_0^\prime d^3v d^3v^\prime \rang + \text{c.c.},
\end{eqnarray}
where we separate $\Dphi = K_\| + K_d$ and `c.c.' denotes complex conjutage.  These can be re-expressed in terms of linear operators by writing them in the form
\begin{eqnarray}
    D = (\tilde{g}, {\cal D}\tilde{g}) = \lang T_i\int d^3v\frac{\tilde{g}^*}{F_0} {\cal D}\tilde{g} \rang,\\
    K_\| = (\tilde{g}, {\cal K}_\|\tilde{g}) = \lang T_i\int d^3v\frac{\tilde{g}^*}{F_0} {\cal K}_\|\tilde{g} \rang,\\
    K_d = (\tilde{g}, {\cal K}_d\tilde{g}) = \lang T_i\int d^3v\frac{\tilde{g}^*}{F_0} {\cal K}_d\tilde{g} \rang.
\end{eqnarray}
Identifying ${\cal D}$ and ${\cal K}_d$ is simply a matter of exchanging labels of dummy variables of integration ($v$ for $v^\prime$, {\em etc.}).  Manipulating the expression for $K_\|$ to reveal ${\cal K}_\|$ is more involved.  We also need to integrate by parts in $l$ and will need to use the fact that $\partial/\partial l$ is performed at fixed phase space variables $E_i$ and $\mu = m_i \vperp^2/(2B)$.  The velocity space volume element contains an important factor of $1/v_\|$, which generally depends on $l$ and does not itself commute with $\partial/\partial l$:

\begin{equation}
    d^3v = 2 \pi v_\perp dv_\perp v_\| = \sum_\sigma \frac{2 \pi B dE_i d\mu_i}{m_i^2 | v_\| |}
\end{equation}
where $\sigma$ denotes the sign of $v_\|$.

\section{Moment form of eigenproblem}\label{moment-forms-appx}
The three terms of Eqn.~\ref{kinetic-eigenproblem} can be rewritten in terms of the moments of $\tilde{g}$ (Eqn.~\ref{moments-def}):

\begin{eqnarray}
    {\cal D} \tilde{g} = \frac{i\lambda}{2}\osT J_0F_0\left[ \frac{v^2}{\vth^2}\kappa_1 - \kappa_2\right],\\
    {\cal K}_\| \tilde{g} = \frac{\lambda}{2}F_0\left[ \vth J_0\left(-B\frac{\partial}{\partial l}\left(\frac{\kappa_4}{B}\right) + \kappa_5 \right) + \frac{v_\|}{\lambda} \frac{\partial}{\partial l}\left(J_0 \lambda \kappa_1\right)\right],\\
    {\cal K}_d \tilde{g} = \frac{i\lambda}{2} \od J_0F_0\left[ \left(\frac{v_\perp^2}{2\vth^2} + \frac{v_\|^2}{\vth^2}\right)\kappa_1 - \kappa_3\right],\\
\end{eqnarray}
Then, taking moments of Eqn.~\ref{kinetic-eigenproblem} yields the following five equations
\begin{eqnarray}
\frac{2\Lambda}{\lambda} \kappa_1 =i\osT\left(G_1\kappa_1 - G_0\kappa_2\right) - i\Delta\od\left(G_3\kappa_1 - G_0\kappa_3\right) -\Delta G_0\vth \left(\kappa_5 - B\frac{\partial}{\partial l}\left(\frac{\kappa_4}{B}\right)\right),\\
\frac{2\Lambda}{\lambda} \kappa_2 =i\osT\left(G_2\kappa_1 - G_1\kappa_2\right) - i\Delta\od\left( G_2 \kappa_1 - G_1 \kappa_3 \right) -\Delta G_1\vth \left(\kappa_5 - B\frac{\partial}{\partial l}\left(\frac{\kappa_4}{B}\right)\right),\\
\frac{2\Lambda}{\lambda} \kappa_3 =i\osT\left(G_4\kappa_1 - G_3\kappa_2\right) - i\Delta\od\left( G_5 \kappa_1 - G_3\kappa_3 \right) -\Delta G_3\vth \left(\kappa_5 - B\frac{\partial}{\partial l}\left(\frac{\kappa_4}{B}\right)\right),\\
\frac{2\Lambda}{\lambda} \kappa_4 = -\Delta\vth\left( G_{0,2}^\prime\kappa_1 + \frac{G_{0,2}}{\lambda}\frac{\partial }{\partial l}(\lambda \kappa_1)\right),\\
\frac{2\Lambda}{\lambda} \kappa_5 = -\Delta\vth\left( G_{0,2}^{\prime\prime}\kappa_1 + \frac{G_{0,2}^\prime}{\lambda}\frac{\partial}{\partial l}(\lambda \kappa_1)\right).
\end{eqnarray}
See the next section where the integrals $G_{m,n}$, {\em etc.}, are defined and evaluated.  Note that the final two equations can be immediately used to eliminate $\kappa_4$ and $\kappa_5$, leaving a system of three equations.  The second and third equations are used together to find forms for $\kappa_2$ and $\kappa_3$ in terms of $\kappa_1$, and these forms are substituted into the first equation to obtain the final form, in terms of $\kappa_1$ only, given by Eqn.~\ref{kappa1-ode-eqn}.

\section{Bessel-type integrals}\label{bessel-integrals-appx}

The following definitions, mostly copied from \cite{plunk_helander_2022}, are needed to perform the various integrals that appear in the moment equations for our eiegenproblem.  First we need a general form of Weber's integral \citep{encyclopedia-mathematics},

\begin{align}
    {\cal I}_{n}(p,a_1,a_2) &= \int_0^\infty \exp(-p t^2) J_{n}(a_1 t)J_{n}(a_2 t) t dt\nonumber\\
    &= \frac{1}{2p} \exp\left( \frac{-a_1^2-a_2^2}{4p}\right) I_{n}\left(\frac{a_1 a_2}{2p}\right)
\end{align}
where $I_{n}$ is the modified Bessel function of order $n$.  The integrals we need to evaluate can be conveniently found in terms of ${\cal I}_{n}$.  We define

\begin{subeqnarray}
    G_{\perp m}(b) &= &2 \int_0^{\infty} x_\perp^{m+1} \exp(-x_\perp^2) J_0^2(\sqrt{2 b} x_\perp) dx_\perp,\\
    G_{\perp m}^{(1)}(b) &= &2 \int_0^{\infty} x_\perp^{m+2} \exp(-x_\perp^2) J_0(\sqrt{2 b} x_\perp)J_1(\sqrt{2 b} x_\perp) dx_\perp,\\
    G_{\perp m}^{(2)}(b) &= &2 \int_0^{\infty} x_\perp^{m+3} \exp(-x_\perp^2) J_1^2(\sqrt{2 b} x_\perp) dx_\perp,
\end{subeqnarray}
where $m$ is assumed to be even.  Now we note that these integrals can be evaluated in terms of Weber's integral:

\begin{subeqnarray}\label{eq:Gperp}
    G_{\perp m}(b) &= &2\left[ \left(-\frac{d}{dp}\right)^{m/2} {\cal I}_0(p,\sqrt{2b},\sqrt{2b})\right]_{p=1},\\
    G_{\perp m}^{(1)}(b) &= &2\left[ \left(-\frac{d}{dp}\right)^{m/2} \left(-\frac{d}{d\lambda}\right) {\cal I}_0(p,\lambda,\sqrt{2b})\right]_{p=1, \lambda = \sqrt{2b}},\\
    G_{\perp m}^{(2)}(b) &= &2\left[ \left(-\frac{d}{dp}\right)^{m/2}\left(-\frac{d}{d\lambda_1}\right)\left(-\frac{d}{d\lambda_2}\right) {\cal I}_0(p,\lambda_1,\lambda_2)\right]_{p=1, \lambda_1 = \lambda_2 = \sqrt{2b}}.
\end{subeqnarray}
The above relations allows us to evaluate the functions

\begin{subeqnarray}\label{eq:Gmn}
    G_{m,n}(b) &= &G_{\perp m}(b) G_{\parallel n},\\
    G_{m,n}^{(1)}(b) &= &G_{\perp m}^{(1)}(b) G_{\parallel n},\\
    G_{m,n}^{(2)}(b) &= &G_{\perp m}^{(2)}(b) G_{\parallel n}.
\end{subeqnarray}
where 

\begin{equation}
    G_{\parallel n} = \frac{1}{\sqrt{\pi}}\int_{-\infty}^{\infty} \exp(-x_\parallel^2) x_\parallel^n dx_\parallel = \frac{1+(-1)^n}{2\sqrt{\pi}} \Gamma_E\left(\frac{1+n}{2}\right),
\end{equation}
and $\Gamma_E$ is the gamma function.  Finally, we can evaluate the integrals $G_{m,n}^{\prime}$, and $G_{m,n}^{\prime}$.  We define

\begin{eqnarray}
G_{\perp m}^{\prime}(b) &= &2 \int_0^{\infty} x_\perp^{m+1} \exp(-x_\perp^2) J_0 \frac{\partial J_0}{\partial l} dx_\perp,\\
G_{\perp m}^{\prime\prime}(b) &= &2 \int_0^{\infty} x_\perp^{m+1} \exp(-x_\perp^2) \left(\frac{\partial J_0}{\partial l}\right)^2 dx_\perp,
\end{eqnarray}
Relating $x_\perp$ to the proper gyrokinetic phase space variable $\mu$, that is $x_\perp = (\mu B/T_i)^{1/2}$ allows the derivatives to be evaluated

\begin{equation}
    \frac{\partial J_0}{\partial l} = -\frac{1}{2} x_\perp\sqrt{2b} J_1\frac{\partial}{\partial l}\ln(b B)
\end{equation}

so that we can write

\begin{eqnarray}
G_{\perp m}^{\prime}(b) &= &-\sqrt{b/2} \left(\frac{\partial}{\partial l}\ln(b B) \right) G_{\perp m}^{(1)}(b),\label{eq:GperpPrime}\\
G_{\perp m}^{\prime\prime}(b) &= & b/2 \left(\frac{\partial}{\partial l}\ln(b B)\right)^2 G_{\perp m}^{(2)}(b).
\end{eqnarray}
These expressions allow us to evaluate

\begin{eqnarray}\label{Gprime-eqns}
    G_{m,n}^\prime = G_{\perp m}^{\prime} G_{\parallel n},\\
    G_{m,n}^{\prime\prime} = G_{\perp m}^{\prime\prime} G_{\parallel n}.\label{eq:GPrimePrime}
\end{eqnarray}

\subsection{Explicit expressions for some Bessel integrals}\label{bessel-integrals-appx-b}

The $b(l)$-dependent factors in Eqn.~\ref{kappa1-ode-eqn} can be written as
\begin{eqnarray}
G_{0} &= &G_{0,0},\\
G_{1} &= &G_{2,0} + G_{0,2},\\
G_{2} &= &G_{4,0} + 2 G_{2,2} + G_{0,4},\\
G_{3} &= &\frac{1}{2}G_{2,0} + G_{0,2},\\
G_{4} &= &\frac{1}{2}G_{4,0} + \frac{3}{2} G_{2,2} + G_{0,4},\\
G_{5} &= &\frac{1}{4}G_{4,0} + G_{2,2} + G_{0,4},
\end{eqnarray}
which can be evaluated using the identities of the previous section in terms of the familiar $\Gamma_n(b)$ of gyrokinetic theory (suppressing its argument for succinctness):

\begin{eqnarray}
G_{0} &= &\Gamma_0,\\
G_{1} &= &\left(\frac{3}{2}-b\right) \Gamma_0+b \Gamma_1,\\
G_{2} &= &\frac{1}{4} \left(\left(6 b^2-20 b+15\right) \Gamma_0 + 2 b\left((10-4 b) \Gamma_1 + b \Gamma_2\right)\right),\\
G_{3} &= &\frac{1}{2} \left(b \Gamma_1-(b-2) \Gamma_0\right),\\
G_{4} &= &\frac{1}{4} \left(\left(3 b^2-11 b+10\right) \Gamma_0+b \left((11-4 b) \Gamma_1+b \Gamma_2\right)\right),\\
G_{5} &= &\frac{1}{8} \left(\left(3 b^2-12 b+14\right) \Gamma_0+b \left(b \Gamma_2-4 (b-3) \Gamma_1\right)\right).
\end{eqnarray}
where we recall

\begin{equation}
    \Gamma_n(b) = \exp(-b)I_n(b)
\end{equation}

For completeness, we evaluate the few remaining factors that enter Eqn.~\ref{kappa1-ode-eqn}.

\begin{eqnarray}
G_{0,2} = \frac{\Gamma _0}{2},\\
G_{0,2}^{(1)} = \frac{\sqrt{b} \left(\Gamma _0-\Gamma _1\right)}{2 \sqrt{2}},\\
G_{0,2}^{(2)} = \frac{1}{8} \left(3 b \Gamma _0+(2-4 b) \Gamma _1+b \Gamma _2\right),
\end{eqnarray}
and using Eqn.~\ref{Gprime-eqns}

\begin{eqnarray}
    G_{0,2}^\prime = -\sqrt{b/2} \frac{\partial}{\partial l}\left(\ln(b B)\right)\; G_{0,2}^{(1)},\\
    G_{0,2}^{\prime\prime} = b/2 \left(\frac{\partial}{\partial l}\ln(b B)\right)^2\; G_{0,2}^{(2)}.
\end{eqnarray}

\subsection{Limit $b \rightarrow 0$}\label{small-b-limit-appx}
In the limit $b \rightarrow 0$ we obtain

\begin{eqnarray}
G_{0} &= &1,\\
G_{1} &= &\frac{3}{2},\\
G_{2} &= &\frac{15}{4},\\
G_{3} &= &1,\\
G_{4} &= &\frac{5}{2},\\
G_{5} &= &\frac{7}{4}.
\end{eqnarray}
and

\begin{eqnarray}
G_{0,2} = \frac{1}{2},\\
G_{0,2}^{(1)} = 0,\\
G_{0,2}^{(2)} = 0,
\end{eqnarray}



\section{Linear dispersion relation}\label{local-disp-appx}

For comparison the following local linear dispersion relation can be solved numerically (here we neglect finite Larmor radius effects $J_0 = 1$, and assume zero density gradient)

\begin{equation}
1 + \tau = \frac{2}{\sqrt{\pi}}\int_0^{\infty} x_\perp d x_\perp\int_{-\infty}^{\infty} dx_\| \left[\frac{\Omega - (x_\perp^2 + x_\|^2 - 3/2) }{\Omega - \kappa_d^{-1}(x_\|^2 + x_\perp^2/2) - x_\| \kappa_\|^{-1}} \right]\exp(-x_\|^2-x_\perp^2),\label{eq:local-disp-reln}
\end{equation}
where $\Omega = \omega/(\omega_* \eta)$.  The velocity integrals can be evaluated separately for the $k_\| = 0$ and $\omega_d = 0$ cases in terms of the plasma dispersion function \citep{kadomtsev-pogutse, biglari}.

\bibliographystyle{unsrtnat}
\bibliography{energy-bounds-part-3}

\begin{thebibliography}{15}
\providecommand{\natexlab}[1]{#1}
\providecommand{\url}[1]{\texttt{#1}}
\expandafter\ifx\csname urlstyle\endcsname\relax
  \providecommand{\doi}[1]{doi: #1}\else
  \providecommand{\doi}{doi: \begingroup \urlstyle{rm}\Url}\fi

\bibitem[Helander and Plunk(2022)]{helander_plunk_2022}
P.~Helander and G.G. Plunk.
\newblock Energetic bounds on gyrokinetic instabilities. {P}art 1.
  {F}undamentals.
\newblock \emph{Journal of Plasma Physics}, 88\penalty0 (2):\penalty0
  905880207, 2022.
\newblock \doi{10.1017/S0022377822000277}.

\bibitem[Plunk and Helander(2022)]{plunk_helander_2022}
G.G. Plunk and P.~Helander.
\newblock Energetic bounds on gyrokinetic instabilities. {P}art 2. {M}odes of
  optimal growth.
\newblock \emph{Journal of Plasma Physics}, 88\penalty0 (3):\penalty0
  905880313, 2022.
\newblock \doi{10.1017/S0022377822000496}.

\bibitem[Landreman et~al.(2015)Landreman, Plunk, and
  Dorland]{landreman_plunk_dorland_2015}
Matt Landreman, Gabriel~G. Plunk, and William Dorland.
\newblock Generalized universal instability: transient linear amplification and
  subcritical turbulence.
\newblock \emph{Journal of Plasma Physics}, 81\penalty0 (5):\penalty0
  905810501, 2015.
\newblock \doi{10.1017/S0022377815000495}.

\bibitem[Dorland and Hammett(1993)]{dorland-hammett-POP-1993}
W.~Dorland and G.~W. Hammett.
\newblock Gyrofluid turbulence models with kinetic effects.
\newblock \emph{Physics of Fluids B: Plasma Physics}, 5\penalty0 (3):\penalty0
  812--835, 1993.
\newblock \doi{10.1063/1.860934}.
\newblock URL \url{https://doi.org/10.1063/1.860934}.

\bibitem[Plunk et~al.(2014)Plunk, Helander, Xanthopoulos, and
  Connor]{plunk-POP-2014}
G.~G. Plunk, P.~Helander, P.~Xanthopoulos, and J.~W. Connor.
\newblock Collisionless microinstabilities in stellarators. {III}. the
  ion-temperature-gradient mode.
\newblock \emph{Physics of Plasmas}, 21\penalty0 (3):\penalty0 032112, 2014.
\newblock \doi{10.1063/1.4868412}.
\newblock URL \url{https://doi.org/10.1063/1.4868412}.

\bibitem[Schekochihin et~al.(2009)Schekochihin, Cowley, Dorland, Hammett,
  Howes, Quataert, and Tatsuno]{Schekochihin_2009}
A.~A. Schekochihin, S.~C. Cowley, W.~Dorland, G.~W. Hammett, G.~G. Howes,
  E.~Quataert, and T.~Tatsuno.
\newblock Astrophysical gyrokinetics: Kinetic and fluid turbulent cascades in
  magnetized weakly collisional plasmas.
\newblock \emph{The Astrophysical Journal Supplement Series}, 182\penalty0
  (1):\penalty0 310, may 2009.
\newblock \doi{10.1088/0067-0049/182/1/310}.
\newblock URL \url{https://dx.doi.org/10.1088/0067-0049/182/1/310}.

\bibitem[Plunk et~al.(2010)Plunk, Cowley, Schekochihin, and
  Tatsuno]{plunk-JFM-2010}
G.~G. Plunk, S.~C. Cowley, A.~A. Schekochihin, and T.~Tatsuno.
\newblock Two-dimensional gyrokinetic turbulence.
\newblock \emph{Journal of Fluid Mechanics}, 664:\penalty0 407--435, 2010.
\newblock \doi{10.1017/S002211201000371X}.

\bibitem[Helander et~al.(2013)Helander, Proll, and Plunk]{Helander_2013}
P.~Helander, J.~H.~E. Proll, and G.~G. Plunk.
\newblock Collisionless microinstabilities in stellarators. i. analytical
  theory of trapped-particle modes.
\newblock \emph{Physics of Plasmas}, 20\penalty0 (12):\penalty0 122505, 2013.
\newblock \doi{10.1063/1.4846818}.
\newblock URL \url{https://doi.org/10.1063/1.4846818}.

\bibitem[Teaca()]{teaca-priv}
Bogdan Teaca.
\newblock private communication.

\bibitem[Biglari et~al.(1989)Biglari, Diamond, and Rosenbluth]{biglari}
H.~Biglari, P.~H. Diamond, and M.~N. Rosenbluth.
\newblock Toroidal ion-pressure-gradient-driven drift instabilities and
  transport revisited.
\newblock \emph{Physics of Fluids B: Plasma Physics}, 1\penalty0 (1):\penalty0
  109--118, 1989.
\newblock \doi{10.1063/1.859206}.
\newblock URL \url{http://link.aip.org/link/?PFB/1/109/1}.

\bibitem[Helander and Plunk(2021)]{helander-plunk-prl-2021}
P.~Helander and G.~G. Plunk.
\newblock Upper bounds on gyrokinetic instabilities in magnetized plasmas.
\newblock \emph{Phys. Rev. Lett.}, 127:\penalty0 155001, Oct 2021.
\newblock \doi{10.1103/PhysRevLett.127.155001}.
\newblock URL \url{https://link.aps.org/doi/10.1103/PhysRevLett.127.155001}.

\bibitem[Roberg-Clark et~al.(2022{\natexlab{a}})Roberg-Clark, Plunk, and
  Xanthopoulos]{Roberg-Clark-PRR-2022}
G.~T. Roberg-Clark, G.~G. Plunk, and P.~Xanthopoulos.
\newblock Coarse-grained gyrokinetics for the critical ion temperature gradient
  in stellarators.
\newblock \emph{Phys. Rev. Research}, 4:\penalty0 L032028, Aug
  2022{\natexlab{a}}.
\newblock \doi{10.1103/PhysRevResearch.4.L032028}.
\newblock URL \url{https://link.aps.org/doi/10.1103/PhysRevResearch.4.L032028}.

\bibitem[Roberg-Clark et~al.(2022{\natexlab{b}})Roberg-Clark, Xanthopoulos, and
  Plunk]{Roberg-Clark-HSK-2022}
G.~T. Roberg-Clark, P.~Xanthopoulos, and G.~G. Plunk.
\newblock Reduction of electrostatic turbulence in a quasi-helically symmetric
  stellarator via critical gradient optimization, 2022{\natexlab{b}}.
\newblock URL \url{https://arxiv.org/abs/2210.16030}.

\bibitem[Iyanaga and Kawada(1980)]{encyclopedia-mathematics}
S.~Iyanaga and Y.~Kawada.
\newblock \emph{Encyclopedic Dictionary of Mathematics}.
\newblock MIT Press, Cambridge, MA, 1980.

\bibitem[Kadomtsev and Pogutse(1970)]{kadomtsev-pogutse}
B.~B. Kadomtsev and O.~P. Pogutse.
\newblock Turbulence in toroidal systems.
\newblock \emph{Rev. Plasmas Phys.}, 5\penalty0 (6):\penalty0 249--400, 1970.

\end{thebibliography}

\end{document}